# On the Capacity Region of the Two-User Interference Channel

Reza K. Farsani[1]

Email: reza_khosravi@alum.sharif.ir

*Abstract:* **One of the key open problems in network information theory is to obtain the capacity region for the two-user Interference Channel (IC). In this paper, new results are derived for this channel. As a first result, a noisy interference regime is given for the general IC where the sum-rate capacity is achieved by treating interference as noise at the receivers. To obtain this result, a single-letter outer bound in terms of some auxiliary random variables is first established for the sum-rate capacity of the general IC and then those conditions under which this outer bound is reduced to the achievable sum-rate given by the simple treating interference as noise strategy are specified. The main benefit of this approach is that it is applicable for any two-user IC (potentially non-Gaussian). For the special case of Gaussian channel, our result is reduced to the noisy interference regime that was previously obtained. Next, some results are given on the Han-Kobayashi (HK) achievable rate region. The evaluation of this rate region is in general difficult. In this paper, a simple characterization of the HK rate region is derived for some special cases, specifically, for a novel very weak interference regime. As a remarkable characteristic, it is shown that for this very weak interference regime, the achievable sum-rate due to the HK region is identical to the one given by the simple treating interference as noise strategy.**

## I. INTRODUCTION

One of the key open problems in network information is to obtain the capacity region for the two-user Interference Channel (IC). This channel is a simple communication scenario where two transmitters separately send independent messages to their corresponding receivers while causing interference to the other receiver. In recent years, numerous papers have studied capacity limits for this channel and related scenarios. For a comprehensive review of the existing results, please refer to [1] and literature therein. See also [1-7] for recent results.

In this paper, we derive new results for the two-user IC. As a first result, we obtain a noisy interference regime for the general IC (potentially non-Gaussian). Recall that in concurrent papers [8, 9, 10], it was shown that for the Gaussian IC when the interference level is below a certain threshold, the sum-rate capacity is achieved by treating interference as noise at both receivers. A natural question is that how one can adapt this result for the discrete IC? In this paper, by a new approach we re-derive the result of [8, 9, 10] to be applicable not only for the Gaussian channel but also for any two-user IC. In our approach, a single-letter outer bound in terms of some auxiliary random variables is first established for the sum-rate capacity of the general IC and then those conditions under which this outer bound is reduced to the achievable sum-rate due to the simple treating interference as noise strategy are specified.

Next, we present some results on the Han-Kobayashi achievable rate region [11]. This is the best known capacity inner bound for the two-user IC which was derived in 1981. A compact characterization of the HK rate region was derived in [12]. However, the evaluation of the HK rate region for the general case is still difficult due to numerous degrees of freedom involved in the problem [8]. In this paper, we propose a new decoding scheme for the two-user IC named the *semi-joint decoding* and demonstrate that, despite its simple structure, it has the same performance as the HK scheme for a novel very weak interference regime and also for the strong interference regime. Consequently, we derive a simple characterization of the HK region for the identified very weak interference regime. As a remarkable characteristic, we also show that for this regime, the achievable sum-rate due to the HK region is identical to the one given by the simple treating interference as noise strategy. We simplify the HK region for the IC with strong interference at one receiver and the one-sided IC, as well.

In the following section, we present the channel model. Our main results are given in section III.

---

[1] Reza K. Farsani was with the department of electrical engineering, Sharif University of Technology. He is by now with the school of cognitive sciences, Institute for Research in Fundamental Sciences (IPM), Tehran, Iran.

## II. CHANNEL MODEL

In this paper, we use the same notations as those in [1], most of them are by now standard.

*The Two-User IC:* In this channel, two transmitters send independent messages to their respective receivers via a common media. Figure 1 depicts the channel model.

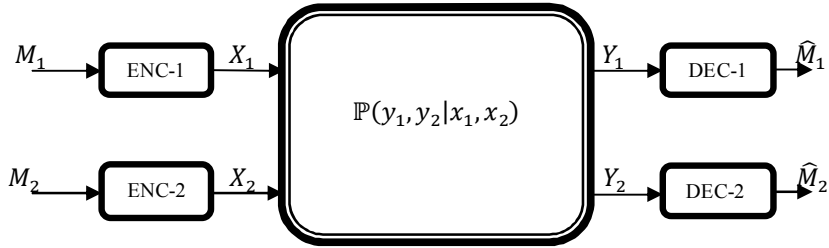

Figure 1. The two-user IC: The transmitter $X_i$ sends the message $M_i$ to its respective receiver $Y_i$, $i = 1,2$.

The two-user IC is given by the input signals $X_1 \in \mathcal{X}_1$ and $X_2 \in \mathcal{X}_2$, the outputs $Y_1 \in \mathcal{Y}_1$ and $Y_2 \in \mathcal{Y}_2$ and the transition probability function $\mathbb{P}(y_1, y_2 | x_1, x_2)$ that describes the relation between the inputs and the outputs of the channel. The channel is assumed to be memoryless. The Gaussian IC is usually formulated in the following standard form:

$$\begin{cases} Y_1 = X_1 + aX_2 + Z_1 \\ Y_2 = bX_1 + X_2 + Z_2 \end{cases} \tag{1}$$

where $Z_1, Z_2$ are zero-mean unit-variance Gaussian random variables (RVs), $a$ and $b$ are fixed real numbers, and $\mathbb{E}[X_i^2] \leq P_i$, $i = 1,2$. A two-user IC is said to be one-sided if either one of the following conditions holds:

$$\begin{cases} \mathbb{P}(y_1, y_2 | x_1, x_2) = \mathbb{P}(y_1 | x_1, x_2) \mathbb{P}(y_2 | x_2) & (a) \\ \mathbb{P}(y_1, y_2 | x_1, x_2) = \mathbb{P}(y_1 | x_1) \mathbb{P}(y_2 | x_1, x_2) & (b) \end{cases} \tag{2}$$

In other words, for a one-sided IC, the signal of one receiver does not depend on the signal of the non-respective transmitter. The Gaussian one-sided IC is also derived by setting either $a = 0$ or $b = 0$ in (2). In this paper, without loss of generality, we only consider the one-sided ICs that satisfy the condition (2-a).

## III. MAIN RESULTS

### III.A ) A Noisy Interference Regime for the General IC

Our first result is to identify a noisy interference regime for any two-user IC. Let us first present a new lemma which is critical in our future derivation.

***Lemma 1)*** *Given an integer $n$, let $Y_1^n$ and $Y_2^n$ be two arbitrary random vector of length $n$. For any arbitrary random variable A, we have:*

$$H(Y_1^n | A) - H(Y_2^n | A) = \sum_{t=1}^{n} \left( H(Y_{1,t} | U_t, A) - H(Y_{2,t} | U_t, A) \right) \tag{3}$$

where $U_t = (Y_1^{t-1}, Y_{2,t+1}^n)$, $t = 1,2, \ldots, n$.

Reza K. Farsani, 2013

*Proof of Lemma 1)* We can write:

$$H(Y_1^n|A) - H(Y_2^n|A) = \sum_{t=1}^n \left( H(Y_{1,t}|Y_1^{t-1},A) - H(Y_{2,t}|Y_{2,t+1}^n,A) \right)$$

$$= \sum_{t=1}^n \left( I(Y_{2,t+1}^n; Y_{1,t}|Y_1^{t-1},A) + I(Y_{2,t+1}^n, A; Y_{2,t}) \right) + \sum_{t=1}^n \left( H(Y_{1,t}|Y_1^{t-1}, Y_{2,t+1}^n, A) - H(Y_{2,t}) \right)$$

$$= \sum_{t=1}^n \left( I(Y_{2,t+1}^n; Y_{1,t}|Y_1^{t-1},A) + I(Y_1^{t-1}, Y_{2,t+1}^n, A; Y_{2,t}) - I(Y_1^{t-1}; Y_{2,t}|Y_{2,t+1}^n, A) \right) + \sum_{t=1}^n \left( H(Y_{1,t}|Y_1^{t-1}, Y_{2,t+1}^n, A) - H(Y_{2,t}) \right)$$

$$\stackrel{(a)}{=} \sum_{t=1}^n I(Y_1^{t-1}, Y_{2,t+1}^n, A; Y_{2,t}) + \sum_{t=1}^n \left( H(Y_{1,t}|Y_1^{t-1}, Y_{2,t+1}^n, A) - H(Y_{2,t}) \right)$$

$$= \sum_{t=1}^n \left( H(Y_{1,t}|Y_1^{t-1}, Y_{2,t+1}^n, A) - H(Y_{2,t}|Y_1^{t-1}, Y_{2,t+1}^n, A) \right) \tag{4}$$

where equality (a) is due to the Csiszar-Korner identity. ∎

In general, the above lemma is a very useful tool for single-letterization while establishing capacity outer bounds for different network scenarios. This lemma holds for both discrete and continuous random variables.

Now, we can derive the main result.

**Theorem 1)** *Consider the two-user IC in Fig. 1. Let* $\widetilde{\mathbb{P}}(\tilde{y}_1, \tilde{y}_2|x_1, x_2)$ *be any virtual memoryless IC with:*

$$\widetilde{\mathbb{P}}(\tilde{y}_1, \tilde{y}_2|x_1, x_2) = \widetilde{\mathbb{P}}(\tilde{y}_1|x_1)\widetilde{\mathbb{P}}(\tilde{y}_2|x_2)$$

*Assume that the following conditions hold:*

$$\begin{cases} I(U; Y_2|X_2, \tilde{Y}_2) \leq I(U; \tilde{Y}_1|X_2, \tilde{Y}_2) \\ I(U; Y_1|X_1, \tilde{Y}_1) \leq I(U; \tilde{Y}_2|X_1, \tilde{Y}_1) \end{cases} \quad \text{for all joint PDFs} \quad P_{X_1} P_{X_2} P_{U|X_1 X_2} \tag{5}$$

*and,*

$$\begin{cases} I(X_1; \tilde{Y}_1|Y_1, Q) = 0 \\ I(X_2; \tilde{Y}_2|Y_2, Q) = 0 \end{cases} \quad \text{for} \quad P_Q^* P_{X_1|Q}^* P_{X_2|Q}^* \tag{6}$$

*where* $P_Q^* P_{X_1|Q}^* P_{X_2|Q}^*$ *is a solution to the following maximization:*

$$\max_{P_Q P_{X_1|Q} P_{X_2|Q}} \left( I(X_1; Y_1, \tilde{Y}_1|Q) + I(X_2; Y_2, \tilde{Y}_2|Q) \right) \tag{7}$$

*The sum-rate capacity is given by:*

$$\max_{P_Q P_{X_1|Q} P_{X_2|Q}} \left( I(X_1; Y_1|Q) + I(X_2; Y_2|Q) \right) \tag{8}$$

*Proof of Theorem 1)* The achievability of (8) is immediately derived by treating interference as noise at the receivers. Let prove the converse part. Consider a length-$n$ code with vanishing average error probability for transmitting the messages $M_1$ and $M_2$ uniformly distributed over the sets $[1:2^{nR_1}]$ and $[1:2^{nR_2}]$, respectively. Define the new random variables $U_{1,t}, U_{2,t}, t = 1, \ldots, n$, as follows:

$$U_{1,t} \triangleq \left( \tilde{Y}_1^{t-1}, Y_{2,t+1}^n, X_2^{t-1}, \tilde{Y}_2^{t-1}, X_{2,t+1}^n, \tilde{Y}_{2,t+1}^n \right)$$

$$U_{2,t} \triangleq \left( \tilde{Y}_2^{t-1}, Y_{1,t+1}^n, X_1^{t-1}, \tilde{Y}_1^{t-1}, X_{1,t+1}^n, \tilde{Y}_{1,t+1}^n \right) \tag{9}$$

Based on the Fano's inequality, we have:

$$n(R_1 + R_2) - \epsilon_n \leq I(X_1^n; Y_1^n) + I(X_2^n; Y_2^n)$$

$$\leq I(X_1^n; Y_1^n, \tilde{Y}_1^n) + I(X_2^n; Y_2^n, \tilde{Y}_2^n)$$

$$= H(\tilde{Y}_1^n) - H(Y_2^n|X_2^n, \tilde{Y}_2^n) + H(\tilde{Y}_2^n) - H(Y_1^n|X_1^n, \tilde{Y}_1^n)$$

$$+ H(Y_1^n|\tilde{Y}_1^n) - H(\tilde{Y}_1^n|X_1^n) + H(Y_2^n|\tilde{Y}_2^n) - H(\tilde{Y}_2^n|X_2^n)$$

$$\stackrel{(a)}{=} H(\tilde{Y}_1^n|X_2^n, \tilde{Y}_2^n) - H(Y_2^n|X_2^n, \tilde{Y}_2^n) + H(\tilde{Y}_2^n|X_1^n, \tilde{Y}_1^n) - H(Y_1^n|X_1^n, \tilde{Y}_1^n)$$

$$+ H(Y_1^n|\tilde{Y}_1^n) - H(\tilde{Y}_1^n|X_1^n) + H(Y_2^n|\tilde{Y}_2^n) - H(\tilde{Y}_2^n|X_2^n)$$

$$\stackrel{(b)}{=} \sum_{t=1}^{n} \left( H(\tilde{Y}_{1,t}|\tilde{Y}_1^{t-1}, Y_{2,t+1}^n, X_2^n, \tilde{Y}_2^n) - H(Y_{2,t}|\tilde{Y}_1^{t-1}, Y_{2,t+1}^n, X_2^n, \tilde{Y}_2^n) \right)$$

$$+ \sum_{t=1}^{n} \left( H(\tilde{Y}_{2,t}|\tilde{Y}_2^{t-1}, Y_{1,t+1}^n, X_1^n, \tilde{Y}_1^n) - H(Y_{1,t}|\tilde{Y}_2^{t-1}, Y_{1,t+1}^n, X_1^n, \tilde{Y}_1^n) \right)$$

$$+ H(Y_1^n|\tilde{Y}_1^n) - \sum_{t=1}^{n} H(\tilde{Y}_{1,t}|X_{1,t}) + H(Y_2^n|\tilde{Y}_2^n) - \sum_{t=1}^{n} H(\tilde{Y}_{2,t}|X_{2,t})$$

$$\leq \sum_{t=1}^{n} \left( H(\tilde{Y}_{1,t}|U_{1,t}, X_{2,t}, \tilde{Y}_{2,t}) - H(Y_{2,t}|U_{1,t}, X_{2,t}, \tilde{Y}_{2,t}) \right)$$

$$+ \sum_{t=1}^{n} \left( H(\tilde{Y}_{2,t}|U_{2,t}, X_{1,t}, \tilde{Y}_{1,t}) - H(Y_{1,t}|U_{2,t}, X_{1,t}, \tilde{Y}_{1,t}) \right)$$

$$+ \sum_{t=1}^{n} H(Y_{1,t}|\tilde{Y}_{1,t}) - \sum_{t=1}^{n} H(\tilde{Y}_{1,t}|X_{1,t}) + \sum_{t=1}^{n} H(Y_{2,t}|\tilde{Y}_{2,t}) - \sum_{t=1}^{n} H(\tilde{Y}_{2,t}|X_{2,t})$$

(10)

where equality (a) holds because $\tilde{Y}_1^n$ is independent of $(X_2^n, \tilde{Y}_2^n)$ and $\tilde{Y}_2^n$ is independent of $(X_1^n, \tilde{Y}_1^n)$, and equality (b) is given by Lemma 1. Applying a standard time-sharing argument to (10), we deduce that the sum-rate capacity of the channel is bounded by:

$$\max_{P_Q P_{X_1|Q} P_{X_2|Q} P_{U_1 U_2|X_1 X_2 Q}} \begin{pmatrix} H(\tilde{Y}_1|U_1, X_2, \tilde{Y}_2, Q) - H(Y_2|U_1, X_2, \tilde{Y}_2, Q) + H(\tilde{Y}_2|U_2, X_1, \tilde{Y}_1, Q) - H(Y_1|U_2, X_1, \tilde{Y}_1, Q) \\ + H(Y_1|\tilde{Y}_1, Q) - H(\tilde{Y}_1|X_1, Q) + H(Y_2|\tilde{Y}_2, Q) - H(\tilde{Y}_2|X_2, Q) \end{pmatrix}$$

(11)

Now, if the conditions (5) hold, we have (note that (5) can readily be extended to include the time-sharing variable):

$$H(\tilde{Y}_1|U_1, X_2, \tilde{Y}_2, Q) - H(Y_2|U_1, X_2, \tilde{Y}_2, Q) \leq H(\tilde{Y}_1|X_2, \tilde{Y}_2, Q) - H(Y_2|X_2, \tilde{Y}_2, Q) = H(\tilde{Y}_1|Q) - H(Y_2|X_2, \tilde{Y}_2, Q)$$

$$H(\tilde{Y}_2|U_2, X_1, \tilde{Y}_1, Q) - H(Y_1|U_2, X_1, \tilde{Y}_1, Q) \leq H(\tilde{Y}_2|X_1, \tilde{Y}_1, Q) - H(Y_1|X_1, \tilde{Y}_1, Q) = H(\tilde{Y}_2|Q) - H(Y_1|X_1, \tilde{Y}_1, Q)$$

(12)

By substituting (12) in (11), we derive that the sum-rate capacity is bounded above by:

$$\max_{P_Q P_{X_1|Q} P_{X_2|Q}} \left( I(X_1; Y_1, \tilde{Y}_1|Q) + I(X_2; Y_2, \tilde{Y}_2|Q) \right)$$

(13)

It is clear that if (6) hold, then (13) is reduced to (8). ∎

Now let consider the Gaussian IC (1). Define a virtual IC as follows:

$$\begin{cases} \tilde{Y}_1 \triangleq b(X_1 + \eta_1 \tilde{Z}_1) \\ \tilde{Y}_2 \triangleq a(X_2 + \eta_2 \tilde{Z}_2) \end{cases}$$

(14)

where $\tilde{Z}_1$ and $\tilde{Z}_2$ are independent Gaussian random variables with zero mean and unit variances. It is assumed that $\tilde{Z}_i$ is correlated to $Z_i$ with correlation coefficient $\rho_i$, $i = 1,2$. One can verify that if the following hold:

$$|b\eta_1| \leq \sqrt{1 - \rho_2^2}, \qquad |a\eta_2| \leq \sqrt{1 - \rho_1^2}$$

(15)

then, the inequalities (5) are satisfied. Moreover, using the entropy power inequality (or the worst case additive noise lemma [13]), one can show that under the conditions (15), Gaussian inputs without time-sharing are the optimal solution for the maximization (7). Accordingly, the conditions (6) are equivalent to the following:

$$\eta_1 \rho_1 = 1 + a^2 P_2, \qquad \eta_2 \rho_2 = 1 + b^2 P_1 \tag{16}$$

As given in [9], if the following holds:

$$|a(1 + b^2 P_1)| + |b(1 + a^2 P_2)| \leq 1 \tag{17}$$

then, the parameters $\eta_1, \eta_2, \rho_1$ and $\rho_2$ can be definitely chosen to satisfy both (15) and (16). Therefore, for the Gaussian IC satisfying (17), the sum-rate capacity is achieved by treating interference as noise at the receivers, as derived in [8, 9, 10]. The difference between our approach and that of [9] is actually in the single-letterization steps. In [9], using the worst case additive noise lemma [13], it is directly shown that the expression on the left side of equation (a) in (10) is bounded by (13) if the conditions (15) are satisfied for the Gaussian IC. But by our approach, using Lemma 1, first a single-letter outer bound in terms of some auxiliary random variables is derived for the general IC. Then, those conditions under which this outer bound is reduced to (13) are specified. The main benefit of our approach is that it is applicable for any two-user IC (potentially non-Gaussian).

*III.B) On Capacity Inner Bounds for the IC*

In this subsection, we present some new results on capacity inner bounds for the two-user IC. The best known achievable rate region for this channel is the well-known HK region [11] in the following:

$$\mathfrak{R}_i^{HK_c} = \bigcup_{\substack{P_Q P_{W_1|Q} P_{W_2|Q} \\ P_{X_1|W_1Q} P_{X_2|W_2Q}}} \begin{cases} (R_1, R_2) \in \mathbb{R}_+^2: \\ R_1 \leq I(X_1; Y_1|W_2, Q) \\ R_2 \leq I(X_2; Y_2|W_1, Q) \\ R_1 + R_2 \leq I(X_1, W_2; Y_1|Q) + I(X_2; Y_2|W_1, W_2, Q) \\ R_1 + R_2 \leq I(X_1; Y_1|W_1, W_2, Q) + I(X_2, W_1; Y_2|Q) \\ R_1 + R_2 \leq I(X_1, W_2; Y_1|W_1, Q) + I(X_2, W_1; Y_2|W_2, Q) \\ 2R_1 + R_2 \leq I(X_1, W_2; Y_1|Q) + I(X_1; Y_1|W_1, W_2, Q) \\ \qquad\qquad + I(X_2, W_1; Y_2|W_2, Q) \\ R_1 + 2R_2 \leq I(X_2; Y_2|W_1, W_2, Q) + I(X_2, W_1; Y_2|Q) \\ \qquad\qquad + I(X_1, W_2; Y_1|W_1, Q) \end{cases} \tag{18}$$

The rate region $\mathfrak{R}_i^{HK_c}$ given in (18) has a compact description compared to the initial HK characterization [11]. This compact characterization is due to [12]. In the initial HK scheme for the two-user IC, each encoder splits its message into two parts named the *common part* and the *private part*, and then these two parts are encoded using independent codewords. The encoders send a deterministic function of these two independent codewords over the channel. By incorporating a jointly decoding technique, each receiver decodes the common part of the messages of both transmitters as well as the private part of its respective transmitter. The rate region due to this simple scheme, if the time-sharing is employed, is equivalent to (18); see [12] for details. The characterization (18), although is a compact version of the initial HK region, its evaluation is still difficult even for the special case of Gaussian channel [8].

In what follows, we present a new decoding scheme for the two-user IC named the semi-joint decoding and discuss its performance versus the HK scheme. Moreover, we simplify the HK region (18) for certain classes of ICs.

First, let us introduce a novel class of ICs with very weak interference.

***Definition:*** *The two-user IC is said to have very weak interference provided that:*

$$I(W_1; Y_2|X_2) \leq I(W_1; Y_1)$$
$$I(W_2; Y_1|X_1) \leq I(W_2; Y_2)$$

(19)

*for all joint PDFs $P_{W_1} P_{W_2} P_{X_1|W_1} P_{X_2|W_2}$.*

By examining the conditions (19) for the Gaussian channel (1) with Gaussian input distributions, we obtain:

$$a^2 \leq \frac{1}{b^2 P_1 + 1}$$
$$b^2 \leq \frac{1}{a^2 P_2 + 1}$$

(20)

These conditions can be viewed as the very weak interference regime for the Gaussian IC (1). Nonetheless, it should be mentioned that (20) is not equivalent to (19), in general. Also, it is worth noting that the regime (20) strictly includes the noisy interference regime (17) as a special case.

In the following theorem, we present our new achievability scheme for the two-user IC. For the very weak interference regime (19), our scheme yields the same achievable rate region as the HK scheme. As a byproduct, the HK region (18) is also simplified for the very weak interference regime (17).

***Theorem 2)*** *Define the rate region $\mathfrak{R}_i^{IC}$ as follows:*

$$\mathfrak{R}_i^{IC} \triangleq \bigcup_{\substack{P_Q P_{W_1|Q} P_{W_2|Q} \\ P_{X_1|W_1 Q} P_{X_2|W_2 Q}}} \begin{Bmatrix} (R_1, R_2) \in \mathbb{R}_+^2: \\ R_1 \leq I(X_1; Y_1|W_2, Q) \\ R_2 \leq I(X_2; Y_2|W_1, Q) \\ R_1 + R_2 \leq I(X_1, W_2; Y_1|Q) + I(X_2; Y_2|W_2, Q) \\ R_1 + R_2 \leq I(X_1; Y_1|W_1, Q) + I(X_2, W_1; Y_2|Q) \end{Bmatrix}$$

(21)

*The set $\mathfrak{R}_i^{IC}$ is an achievable rate region for the two-user IC. Moreover, for any two-user IC satisfying the very weak interference conditions (19), the HK achievable rate region (18) is equivalent to $\mathfrak{R}_i^{IC}$.*

*Proof of Theorem 2)* First, we prove the achievability of (21) for the general IC using a random coding argument in which the superposition encoding technique is exploited at the transmitters and the semi-joint decoding is applied at the receivers. Precisely speaking, each message is split into two parts and then these sub-messages are encoded at transmitters in a superposition fashion. At each receiver, the decoding is performed in two steps:

1. First, the cloud center of the desired message is detected using joint decoding.
2. After decoding the cloud center, the satellite of the desired message is decoded by treating interference (i.e., both the cloud center and the satellite of the non-desired message) as noise.

Note that the parameter $Q$ in the rate region $\mathfrak{R}_i^{IC}$ is a time-sharing RV. Therefore, for simplicity of exposition we derive the achievability of (21) for the case of $Q \equiv \emptyset$. The coding scheme can be readily extended to a time-shared one. Fix the joint PDFs $P_{W_1}(w_1), P_{W_2}(w_2), P_{X_1|W_1}(x_1|w_1)$, and $P_{X_2|W_2}(x_2|w_2)$. For a given $(R_1, R_2) \in \mathbb{R}_+^2$, split the rates $R_1$ and $R_2$ as:

$$R_1 = R_{10} + R_{11}$$
$$R_2 = R_{20} + R_{22}$$

(22)

The messages $M_1$ and $M_2$ are expressed as: $M_i = (M_{i0}, M_{ii}) \in [1:2^{nR_{i0}}] \times [1:2^{nR_{ii}}]$, $i = 1,2$. The random codebook is generated as follows.

*Encoding:* At the $i^{th}$ transmitter, $i = 1,2$, the messages $M_{i0}$ and $M_{ii}$ are encoded in a superposition structure:

1. Generate at random $2^{nR_{i0}}$ i.i.d. codewords $W_i^n$ according to $Pr(w_i^n) = \prod_{t=1}^{n} P_{W_i}(w_{i,t})$. Label these codewords $W_i^n(m_{i0})$, where $m_{i0} \in [1:2^{nR_{i0}}]$.
2. For each $W_i^n(m_{i0})$, where $m_{i0} \in [1:2^{nR_{i0}}]$, generate $2^{nR_{ii}}$ i.i.d. codewords $X_i^n$ according to $Pr(x_i^n) = \prod_{t=1}^{n} P_{X_i|W_i}(x_{i,t}|w_{i,t}(m_{i0}))$. Label these codewords $X_i^n(m_{i0}, m_{ii})$, where $m_{ii} \in [1:2^{nR_{ii}}]$.

This codebook is revealed at both transmitters and also both receivers. Assuming $(M_{i0}, M_{ii}) = (m_{i0}, m_{ii})$, the $i^{th}$ transmitter, $i = 1,2$, sends $X_i^n(m_{i0}, m_{ii})$ over the channel.

*Decoding:* The receivers applied the semi-joint decoding; this type of decoding is performed in two steps, as follows:

At receiver 1, assume that the sequence $Y_1^n$ has been received:

1. At the first step, the receiver tries to find a unique $\hat{m}_{10}$ for which there exist some $\bar{m}_{20}$ and $\bar{m}_{11}$, such that:

$$(W_1^n(\hat{m}_{10}), W_2^n(\bar{m}_{20}), X_1^n(\hat{m}_{10}, \bar{m}_{11}), Y_1^n) \in \mathcal{T}_\epsilon^n(P_{W_1 W_2 X_1 Y_1}) \tag{23}$$

If there exists such $\hat{m}_{10}$, the receiver estimates the transmitted message $m_{10}$ as the corresponding $\hat{m}_{10}$. If there is no such $\hat{m}_{10}$ or there is more than one, the decoder sets $\hat{m}_{10} = 1$ and declares an error. Therefore, in this step the receiver decodes the message $m_{10}$. By a standard analysis of error probability, one can show that this message can be decoded successfully (with arbitrary small probability of error) provided that:

$$R_{10} + R_{11} + R_{20} < I(X_1, W_1, W_2; Y_1) \stackrel{(a)}{=} I(X_1, W_2; Y_1)$$
$$R_{10} + R_{11} < I(X_1, W_1; Y_1|W_2) \stackrel{(b)}{=} I(X_1; Y_1|W_2) \tag{24}$$

where equality (a) and (b) hold because $W_1 \to (X_1, W_2) \to Y_1$ forms a Markov chain.

2. Assuming that decoding at previous step has been performed successfully, at the second step the receiver decodes the transmitted message $m_{11}$ using the sequence $(W_1^n(m_{10}^*), Y_1^n)$, where $m_{10}^*$ is the decoded message at previous step; it tries to find a unique $\hat{m}_{11}$, such that:

$$(W_1^n(m_{10}^*), X_1^n(m_{10}^*, \hat{m}_{11}), Y_1^n) \in \mathcal{T}_\epsilon^n(P_{W_1 X_1 Y_1}) \tag{25}$$

If there is no such $\hat{m}_{11}$ or there is more than one, the receiver sets $\hat{m}_{11} = 1$, and declares an error. The decoding at this step can be performed successfully provided that:

$$R_{11} < I(X_1; Y_1|W_1) \tag{26}$$

The decoding procedure at receiver 2 is similar to receiver 1, expect the indices 1 and 2 should be exchanged everywhere. Therefore, decoding at receiver 2 can be successfully performed provided that:

$$R_{20} + R_{22} + R_{10} < I(X_2, W_1; Y_2)$$
$$R_{20} + R_{22} < I(X_2; Y_2|W_1)$$
$$R_{22} < I(X_2; Y_2|W_2) \tag{27}$$

Now by considering (24), (26), and (27), and then applying a simple Fourier-Motzkin elimination to remove $R_{10}, R_{11}, R_{20}, R_{22}$, we obtain the desired rate region.

Then, we show that the HK rate region (18) under the conditions (19) is equivalent to $\mathfrak{R}_i^{IC}$ in (21). Note that, the conditions (19) can be readily extended as:

$$I(W_1; Y_2|X_2, Q) \leq I(W_1; Y_1|Q)$$
$$I(W_2; Y_1|X_1, Q) \leq I(W_2; Y_2|Q)$$
(28)

Now, suppose $(R_1, R_2) \in \mathfrak{R}_i^{HK_c}$. We have:

$$\begin{aligned}
R_1 + R_2 &\leq I(X_1, W_2; Y_1|W_1, Q) + I(X_2, W_1; Y_2|W_2, Q) \\
&= I(X_1, W_1, W_2; Y_1|Q) - I(W_1; Y_1|Q) + I(X_2; Y_2|W_2, Q) + I(W_1; Y_2|X_2, W_2, Q) \\
&\stackrel{(a)}{=} I(X_1, W_2; Y_1|Q) - I(W_1; Y_1|Q) + I(X_2; Y_2|W_2, Q) + I(W_1; Y_2|X_2, Q) \\
&\stackrel{(b)}{\leq} I(X_1, W_2; Y_1|Q) + I(X_2; Y_2|W_2, Q)
\end{aligned}$$
(29)

where the equality (a) holds because $W_1 \to X_1 W_2 Q \to Y_1$, $W_2 \to X_2 Q \to Y_2$ and also $W_2 \to X_2 W_1 Q \to Y_2$ form Markov chains, and the inequality (b) is due to first condition of (28). Similarly, one can derive:

$$R_1 + R_2 \leq I(X_1; Y_1|W_1, Q) + I(X_2, W_1; Y_2|Q)$$
(30)

Therefore, $(R_1, R_2) \in \mathfrak{R}_i^{IC}$ which demonstrates that $\mathfrak{R}_i^{HK_c} \subseteq \mathfrak{R}_i^{IC}$. Conversely, suppose $(R_1, R_2) \in \mathfrak{R}_i^{IC}$. Then, we have:

$$\begin{aligned}
R_1 + R_2 &\leq I(X_1, W_2; Y_1|Q) + I(X_2; Y_2|W_2, Q) \\
&= I(X_1; Y_1|Q) + I(W_2; Y_1|X_1, Q) + I(X_2; Y_2|W_2, Q) \\
&\stackrel{(a)}{\leq} I(X_1; Y_1|Q) + I(W_2; Y_2|Q) + I(X_2; Y_2|W_2, Q) \\
&= I(X_1; Y_1|Q) + I(X_2; Y_2|Q)
\end{aligned}$$
(31)

where inequality (a) is due to the second condition of (28). Therefore, at least one of the following inequalities holds:

$$R_1 \leq I(X_1; Y_1|Q), \qquad R_2 \leq I(X_2; Y_2|Q)$$
(32)

Without loss of generality assume that: $R_1 \leq I(X_1; Y_1|Q)$. Thereby, $(R_1, R_2)$ belongs to the following subset of $\mathfrak{R}_i^{HK_c}$ which is extracted in consideration of $W_2 \equiv \emptyset$,

$$\bigcup_{P_Q P_{X_1 W_1|Q} P_{X_2|Q}} \begin{cases} (R_1, R_2) \in \mathbb{R}_+^2: \\ R_1 \leq I(X_1; Y_1|Q) \\ R_2 \leq I(X_2; Y_2|W_1, Q) \\ R_1 + R_2 \leq I(X_1; Y_1|W_1, Q) + I(X_2, W_1; Y_2|Q) \end{cases}$$
(33)

which demonstrates that $\mathfrak{R}_i^{IC} \subseteq \mathfrak{R}_i^{HK_c}$. ∎

***Remarks 1:***

1. It is remarkable to know that the achievable rate region $\mathfrak{R}_i^{IC}$ is optimal in the strong interference regime [14] which yields the capacity region. In fact, the rate region resulting by setting $W_1 \equiv X_1$ and $W_2 \equiv X_2$ in (21) is identical to the capacity region of the channel with strong interference [14]. This demonstrates that the semi-joint decoding scheme is useful not only in the very weak interference regime (19) but also in the strong interference regime.
2. For the Gaussian IC (1) in the very weak interference regime (19), the HK rate region (18) and the region (21) for Gaussian input distributions without time-sharing are equivalent. On the one hand, for the HK scheme in the weak interference regime ($|a| \leq 1$ and $|b| \leq 1$), time-sharing and concavification result in the same region [8, Th. 7]. Therefore, in the regime (19), the two regions (18) and (21) for Gaussian input distributions with time-sharing are also equivalent.

Using Theorem 2, one can deduce the following interesting result for the ICs in the very weak interference regime (19).

***Corollary 1)*** *For the two-user discrete IC with very weak interference (19), the maximum achievable sum-rate due to the HK region (18) is given by:*

$$\max_{P_{X_1} P_{X_2}} \left( I(X_1; Y_1) + I(X_2; Y_2) \right)$$

(34)

*Proof of Corollary 1)* According to Theorem 2, in the very weak interference regime (19), the HK region is equivalent to $\Re_i^{IC}$ in (21). For the region $\Re_i^{IC}$ in (21), the sum-rate is bounded as:

$$\begin{aligned}
R_1 + R_2 &\leq I(X_1, W_2; Y_1|Q) + I(X_2; Y_2|W_2, Q) \\
&= I(X_1; Y_1|Q) + I(W_2; Y_1|X_1, Q) + I(X_2; Y_2|W_2, Q) \\
&\stackrel{(a)}{\leq} I(X_1; Y_1|Q) + I(W_2; Y_2|Q) + I(X_2; Y_2|W_2, Q) \\
&= I(X_1; Y_1|Q) + I(X_2; Y_2|Q) \\
&\leq \max_q \left( \left( I(X_1; Y_1|Q=q) + I(X_2; Y_2|Q=q) \right) \right) \\
&\leq \max_{P_{X_1} P_{X_2}} \left( I(X_1; Y_1) + I(X_2; Y_2) \right)
\end{aligned}$$

(35)

where inequality (a) is due to the second condition of (28). ∎

***Remark 2:*** Corollary 1 indicates that in the very weak interference regime (19), the maximum achievable sum-rate due to the HK region is identical to the one given by the simple treating interference as noise strategy.

Next, we consider a two-user IC with strong interference at one receiver. Specifically, let examine the case where the second receiver perceives strong interference. For such a channel, the following holds:

$$I(X_1; Y_1|X_2) \leq I(X_1; Y_2|X_2) \quad \text{for all joint PDFs} \quad P_{X_1}(x_1) P_{X_2}(x_2)$$

(36)

In the next theorem, we simplify the HK rate region (18) for this channel.

***Theorem 3)*** *Define the rate region $\Re_i^{HKSI \to Y_2}$ as follows:*

$$\Re_i^{HKSI \to Y_2} \triangleq \bigcup_{P_Q P_{X_1 W_1|Q} P_{X_2|Q}} \left\{ \begin{array}{l} (R_1, R_2) \in \mathbb{R}_+^2 : \\ R_1 \leq I(X_1; Y_1|W_2, Q) \\ R_2 \leq I(X_2; Y_2|X_1, Q) \\ R_1 + R_2 \leq I(X_1, X_2; Y_2|Q) \\ R_1 + R_2 \leq I(X_1, W_2; Y_1|Q) + I(X_2; Y_2|X_1, W_2, Q) \\ 2R_1 + R_2 \leq I(X_1, W_2; Y_1|Q) + I(X_1, X_2; Y_2|W_2, Q) \end{array} \right\}$$

(37)

*For any two-user IC with strong interference at the receiver $Y_2$, where the condition (36) is satisfied, the HK achievable rate region (18) is equivalent to $\Re_i^{HKSI \to Y_2}$.*

*Proof of Theorem 3)* First, suppose that $(R_1, R_2) \in \Re_i^{HK_c}$. First note that the condition (36) can be extended as follows (see [3, Sec. II, Lemma 1]):

$$I(X_1; Y_1|X_2, D) \leq I(X_1; Y_2|X_2, D) \quad \text{for all joint PDFs} \quad P_{DX_1X_2}(d, x_1, x_2)$$

(38)

Therefore, we have:

$$I(X_1; Y_1|X_2, W_1, Q) \leq I(X_1; Y_2|X_2, W_1, Q)$$
$$I(X_1; Y_1|X_2, W_2, Q) \leq I(X_1; Y_2|X_2, W_2, Q)$$

(39)

for all joint PDFs $P_{QW_1W_2X_1X_2}(q, w_1, w_2, x_1, x_2)$, specifically, for the joint PDFs that correspond to the HK rate region (18). Also, in the following, we repeatedly use the following Markov relations:

$$W_1, W_2, Q \to X_1, X_2 \to Y_1, Y_2$$
$$W_1 \to X_1, W_2, Q \to Y_1, Y_2$$
$$W_2 \to X_2, W_1, Q \to Y_1, Y_2$$
$$W_1 \to X_1, Q \to Y_1, Y_2$$
$$W_2 \to X_2, Q \to Y_1, Y_2$$

(40)

Now, we can write:

$$\begin{aligned}R_2 &\leq I(X_2; Y_2|W_1, Q) \\ &\overset{(a)}{\leq} I(X_2; Y_2|X_1, W_1, Q) \\ &= I(X_2; Y_2|X_1, Q)\end{aligned}$$

(41)

where inequality (a) holds because $X_2$ is independent of $X_1, W_1$, conditioned on $Q$.

$$\begin{aligned}R_1 + R_2 &\leq I(X_1; Y_1|W_1, W_2, Q) + I(X_2, W_1; Y_2|Q) \\ &\overset{(a)}{\leq} I(X_1; Y_1|X_2, W_1, W_2, Q) + I(X_2, W_1; Y_2|Q) \\ &= I(X_1; Y_1|X_2, W_1, Q) + I(X_2, W_1; Y_2|Q) \\ &\leq I(X_1; Y_2|X_2, W_1, Q) + I(X_2, W_1; Y_2|Q) \\ &= I(X_1, X_2; Y_2|Q)\end{aligned}$$

(42)

where inequality (a) holds because $X_1, W_1$ are independent of $X_2, W_2$, conditioned on $Q$.

$$\begin{aligned}R_1 + R_2 &\leq I(X_1, W_2; Y_1|Q) + I(X_2; Y_2|W_1, W_2, Q) \\ &\overset{(a)}{\leq} I(X_1, W_2; Y_1|Q) + I(X_2; Y_2|X_1, W_1, W_2, Q) \\ &= I(X_1, W_2; Y_1|Q) + I(X_2; Y_2|X_1, W_2, Q)\end{aligned}$$

(43)

where inequality (a) holds because $X_2, W_2$ are independent of $X_1, W_1$, conditioned on $Q$.

$$\begin{aligned}2R_1 + R_2 &\leq I(X_1, W_2; Y_1|Q) + I(X_1; Y_1|W_1, W_2, Q) + I(X_2, W_1; Y_2|W_2, Q) \\ &\leq I(X_1, W_2; Y_1|Q) + I(X_1; Y_1|X_2, W_1, W_2, Q) + I(X_2, W_1; Y_2|W_2, Q) \\ &\leq I(X_1, W_2; Y_1|Q) + I(X_1; Y_2|X_2, W_1, W_2, Q) + I(X_2, W_1; Y_2|W_2, Q) \\ &= I(X_1, W_2; Y_1|Q) + I(X_1, X_2; Y_2|W_2, Q)\end{aligned}$$

(44)

Therefore, $(R_1, R_2) \in \mathfrak{R}_i^{HK_c}$, which demonstrates that $\mathfrak{R}_i^{HK_c} \subseteq \mathfrak{R}_i^{HK_{SI \to Y_2}}$. Conversely, suppose that $(R_1, R_2) \in \mathfrak{R}_i^{HK_{SI \to Y_2}}$. We consider two different cases:

*Case 1)* $R_1, R_2$ also satisfy both the following constraints:

$$R_2 \leq I(X_2; Y_2|X_1, W_2, Q) + I(W_2; Y_1|X_1, Q)$$
$$R_1 + R_2 \leq I(X_1, X_2; Y_2|W_2, Q) + I(W_2; Y_1|X_1, Q)$$

(45)

In this case, $(R_1, R_2)$ belongs to the rate region $\mathfrak{R}_i^{HK_c}$ in consideration of $W_1 \equiv X_1$.

*Case 2)* $R_1, R_2$ violate at least one of the constraints in (45). In this case, one can verify that $R_1 \leq I(X_1; Y_1|Q)$. Thereby, $(R_1, R_2)$ belongs to the rate region $\mathfrak{R}_i^{HK_c}$ in consideration of $W_1 \equiv X_1$ and $W_2 \equiv \emptyset$. This completes the proof. ∎

Let consider the Gaussian IC (1) with mixed interference $|a| \leq 1 \leq |b|$ where the condition (36) is satisfied. For this channel, the HK achievable rate region (18) for Gaussian input distributions was explicitly characterized in [8, Sec. VI.C]. One can readily verify that the characterization (37) also leads to that of [8, Sec. VI.C].

Finally, we simplify the HK rate region for the two-user one-sided IC in (2-a).

**Proposition 1)** For the two-user one-sided IC in (2-a), the HK achievable rate region (18) is equivalent to the following:

$$\bigcup_{P_Q P_{X_1|Q} P_{X_2 W_2|Q}} \begin{Bmatrix} (R_1, R_2) \in \mathbb{R}_+^2: \\ R_1 \leq I(X_1; Y_1|W_2, Q) \\ R_2 \leq I(X_2; Y_2|Q) \\ R_1 + R_2 \leq I(X_1, W_2; Y_1|Q) + I(X_2; Y_2|W_2, Q) \end{Bmatrix}$$

(46)

*Proof of Proposition 1)* First note that for the one-sided IC, the joint PDFs that correspond to the rate region $\mathfrak{R}_i^{HK_c}$ in (18) are given by:

$$P_{QW_1 W_2 X_1 X_2 Y_1 Y_2} = P_Q P_{W_1|Q} P_{W_2|Q} P_{X_1|W_1 Q} P_{X_2|W_2 Q} \mathbb{P}_{Y_1|X_1 X_2} \mathbb{P}_{Y_2|X_2}$$

(47)

Using the distribution (47), it is readily derived that $Y_2$ is indepednt of $W_1$, conditioned on $Q$. By this observation, it is not difficult to show that for the one-sided IC (2-a), the rate region $\mathfrak{R}_i^{HK_c}$ in (18) is optimized in consideration of $W_1 \equiv \emptyset$. Thus, it is equal to (46). ∎

For the special case of the one-sided Gaussian IC, the characterization (46) straightforwardly yields the one derived in [8, Sec. V.C].

Reza K. Farsani, 2013